\documentclass[twocolumn,showpacs,preprintnumbers,amsmath,amssymb]{revtex4}

\usepackage{graphicx}
\usepackage{dcolumn}
\usepackage{bm}

\begin{document}

\title{A nonlinear Ramsey interferometer operating beyond the Heisenberg limit}
\author{S. Choi and B. Sundaram}
\affiliation{Department of Physics, University of Massachusetts,
Boston, MA 02125, USA}

\begin{abstract}
We show that a dynamically evolving two-mode Bose-Einstein
condensate (TBEC) with an adiabatic, time-varying Raman coupling
maps exactly onto a nonlinear Ramsey interferometer that includes a
nonlinear medium. Assuming a realistic quantum state for the TBEC,
namely the $SU(2)$ coherent spin state, we find that the measurement
uncertainty of the ``path-difference'' phase shift scales as the
standard quantum limit ($1/\sqrt{N}$) where $N$ is the number of
atoms, while that for the interatomic scattering strength scales as
$1/N^{7/5}$, overcoming the Heisenberg limit of $1/N$.
\end{abstract}


\maketitle

High-precision interferometry is one of the most important tools of
metrology enabling one to infer various properties of the system
under consideration through the measurement of the phase shift.
Ramsey interferometry provides a way to extract information about
the changes in the system Hamiltonian $H$  at time $t$ via the phase
shift $\phi = \int_{0}^{t} H(t') dt'/\hbar$. In quantum systems, it
is possible to achieve measurement uncertainties approaching the
Heisenberg limit $\Delta \phi \sim 1/N$ where $N$ is the number of
particles conjugate to the phase variable $\phi$ provided one uses
carefully chosen entangled input states such as Schr\"{o}dinger's
Cat state.

It has long been accepted that the Heisenberg limit is the ultimate
limit to measurement; however recently it was shown using parameter
estimation theory that measurement uncertainty of the order $1/N^k$,
where $k$ is the number of parameter-sensitive terms, is
possible\cite{boixo}. In particular, Bose-Einstein condensates
(BECs) with two-body collisions may be able to achieve up to $\Delta
\phi \sim 1/N^{2}$ accuracy in measurement of atom-atom interactions
through a modulation the scattering length using Feshbach resonance
or density variation due to gravitational gradients. In a related
recent work\cite{Lukin}, the quantum limit to the measurement of
atomic scattering length was studied by considering the Heisenberg
exclusion principle applied to a squeezing Hamiltonian, and finding
the optimal spin squeezed state generated using a separate time
dependent Hamiltonian\cite{Unanyan}.

In this paper, we show that the temporal evolution of a TBEC such as
Na atoms in the $|F=1, M_{F} = \pm 1 \rangle$ hyperfine states
trapped in an optical dipole trap with Raman coupling maps directly
onto a nonlinear Ramsey interferometer. A nonlinear interferometer,
as opposed to a normal (linear) interferometer, includes nonlinear
medium in one or both arms. Such TBEC systems have already been
realized experimentally\cite{myatt,stenger}, and as we show, can
achieve measurement accuracy better than the Heisenberg limit. It is
noted that nonlinear interferometers have been studied
previously\cite{nonlinInterf,yurke}, although never in the context
of BEC.

A quantum interferometer can be described in terms of the angular
momentum operators as a transformation operator:
\begin{equation}
\hat{\cal{ I}} = \hat{\cal{ B}}_{-} \hat{\cal{ P}}(\phi) \hat{\cal{
B}}_{+}  = e^{-i\phi \hat{J}_{y}} .
\end{equation}
The 50/50 beam splitter and the phase shifter are given by
$\hat{\cal{ B}}_{\pm} = \exp(\pm i \pi \hat{J}_{x}/2)$ and
$\hat{\cal{ P}}(\phi) = \exp( i \phi \hat{J}_{z})$ where $\hat{J}_x
= \frac{1}{2}
( \hat{J}_{+} + \hat{J}_{-} )$, $\hat{J}_y = \frac{1%
}{2i} ( \hat{J}_{+} - \hat{J}_{-} )$, $\hat{J}%
_{+(-)} = \hat{a}_{1(2)}^{\dagger}\hat{a}_{2(1)}$ and $\hat{J}%
_z = \frac{1}{2} ( \hat{a}_{1}^{\dagger}\hat{a}_{1} -
\hat{a}_{2}^{\dagger}\hat{a}_{2})$ with $\hat{a}_1$ and $%
\hat{a}_{2}$ being the two annihilation operators for the two input
modes into interferometer. For the TBEC with a Raman coupling like
that considered here, the two annihilation operators $\hat{a}_1$ and
$\hat{a}_{2}$ correspond to the atoms in the two hyperfine states.
The time evolution operator $\hat{U}(t)$ for this system defined by
$|\psi(t) \rangle = \hat{U}(t)|\psi(0)\rangle$ is\cite{choi,choi2}:
\begin{equation}
\hat{U}(t) = \hat{R}^{\dagger} e^{-i\hat{H}'t}\hat{R},  \label{U_t}
\end{equation}
where $\hat{R} = e^{-\pi (\hat{J}_{-} - \hat{J}_{+}) /4}$ and $H' = 2 \Omega  \hat{J}_z - \frac{q%
}{2} \hat{J}_{z}^2$. $\Omega$ is the tunneling coupling and $q$ is
the strength of the scattering interaction between the bosons. As
shown earlier\cite{choi}, the detuning of the laser from the
transition between the two species is set to be zero to make the
Hamiltonian diagonal in the $\hat{J}_z$ representation. This also
prevents the generation of an additional geometric phase on top of
the dynamical phase.

The overall action of the time evolution operator $\hat{U}(t)$, Eq.
(\ref{U_t}), can clearly be mapped onto a nonlinear Ramsey
interferometer with the transformation operator
\begin{equation}
\hat{{\cal I}}   =   \hat{{\cal B}}_{-} \hat{{\cal
P}}(\phi^{\prime}_1) \hat{{\cal S}}(\phi^{\prime}_2)  \hat{{\cal
B}}_{+} = e^{-i \phi^{\prime}_1 \hat{J}_x - i \phi^{\prime}_2
\hat{J}_{x}^2/2 }.
\end{equation}
$\hat{{\cal B}}_{-} = \hat{R}^{\dagger} \equiv \exp(-i\pi\hat{J}_{y}
/2)$ and $\hat{{\cal B}}_{+} = \hat{R}$ are the two  50/50 beam
splitters, while $\hat{{\cal P}}(\phi^{\prime}_1) = e^{-i
\phi^{\prime}_1 \hat{J}_z}$ and $\hat{{\cal S}}(\phi^{\prime}_2) =
e^{i \phi^{\prime}_2 \hat{J}_{z}^2/2}$ represent, respectively,  the
``path-difference'' phase shifter and the nonlinear medium. The
phase shifts $\phi^{\prime}_1 = 2\Omega t$ and $\phi^{\prime}_2 = q
t$ are given by the Hamiltonian dynamics. The presence of the
intrinsic temporal evolution has to be taken into account in the
measurement of $\phi^{\prime}_1$ and $\phi^{\prime}_2$ i.e. any
measurement will be shifted at the rate of $2\Omega$ and $q$ per
unit time. Since we are interested in measuring the {\it changes} in
the phases $\phi^{\prime}_i$, $i = 1,2$, we shall redefine
$\phi^{\prime}_i$ to be $\phi^{\prime}_1 = 2\Omega t + \phi_1$ and
$\phi^{\prime}_2 = q t + \phi_2$ and concern ourselves with the
measurement of $\phi_i$.

For our input state we shall consider an $SU(2)$ atomic coherent
state or a coherent spin state (CSS), $| \theta, \phi \rangle$ which
is a reasonable quantum state representing a TBEC\cite{choi2,bloch}.
It is noted that $\hat{R}(t)|\theta ,\phi \rangle = \sum_{m=-j}^{j}
{\cal R}_{m}^{j}(\theta + \lambda, \phi  - \Delta t) |j,m\rangle $
where ${\cal R}_{m}^{j}(\theta, \phi)$ is defined
\begin{eqnarray}
 {\cal R}_{m}^{j}(\theta, \phi) & = &  \left (  \begin{array}{c}   2j  \\ j + m \end{array}  \right )^{1/2} \cos^{j+m} \left ( \frac{\theta}{2} \right )\sin^{j-m} \left ( \frac{\theta }{2} \right )  \nonumber \\
& & \times e^{i(j-m)\phi}. \label{Rjmtp}
\end{eqnarray}
Since the azimuthal angle $\phi$ simply to shifts the origin, we
shall only consider CSS with $\phi = 0$ in this paper. Exotic input
states such as the NOON or the Yurke state\cite{yurke} will be
considered elsewhere as they are currently not yet practical in the
context of TBEC.

The simplest possible scenario is to measure the ``path-difference''
phase shift $\phi_1$ while applying a magnetic field to tune
$\phi'_2 = 0$ via the Feshbach resonance  i.e. no nonlinear
perturbations to the Hamiltonian. This is the standard Ramsey
interferometry which has been studied extensively. The fact that a
TBEC is used instead of the thermal atoms simply provides clean
signals owing to the inherent long range coherence of a condensate.
We will not consider this case any further in this paper. What's
more interesting is the case of finite $q$. Here the presence of the
nonlinear component modifies the interferometric outcome $\phi_1$,
and brings to the fore the question of the uncertainty associated
with measuring the scattering length or $\phi_2$.

First, we analyze TBEC as a nonlinear Ramsey interferometer in the
idealized situation where the measurement of the phase is carried
out as a projective measurement onto a phase state. We estimate the
measurement uncertainty using the Cramers-Rao inequality in such
cases. Then a more practical scheme, measurement of the atom number
difference as a function of the phase shifts $\phi_1$ and $\phi_2$,
is considered along with the corresponding measurement
uncertainties. The fundamental limit to the phase shift measurements
can be calculated by first defining the positive valued operator
measure $\hat{E}(\phi)$ such that the probability density of the
corresponding measurement result is: $P(\phi) = {\rm Tr} [
\hat{\rho} \hat{E}(\phi) ]$ where $\hat{\rho}$ is the density matrix
for the system.  As in Ref. \cite{sanders}, we define the normalized
phase state
\begin{equation}
|j, \Phi \rangle  =  (2j + 1) ^{-1/2} \sum_{m_{x} = -j}^{j} e^{i m_{x} \Phi} |j,m_{x} \rangle_{x}
\end{equation}
so that $\hat{E}(\Phi)d\Phi = (2j + 1)|j, \Phi \rangle \langle j,
\Phi | d\Phi/2\pi$, and for an arbitrary input state $|\psi
\rangle$, the probability density of the measurement result is
\begin{equation}
P(\Phi)  =  \frac{2j + 1}{2\pi }  |\langle \psi | \hat{{\cal
I}}^{\dagger} |j, \Phi \rangle |^2.
\end{equation}

\begin{figure}
\begin{center}
\centerline{\includegraphics[height=7cm]{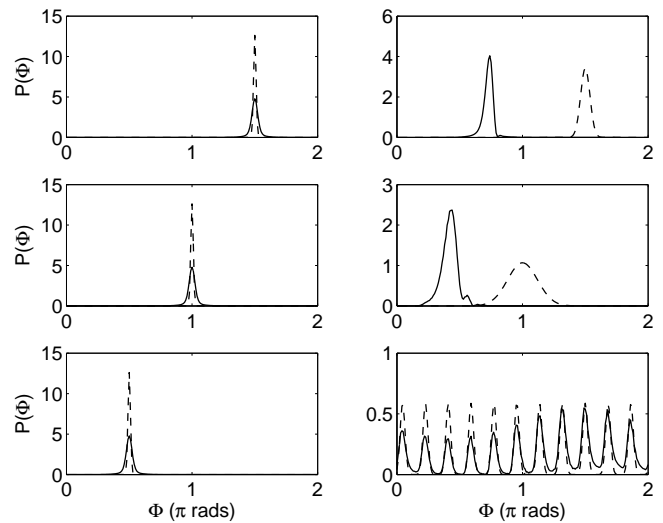}}
\caption{Probability density $P(\Phi)$ for the initial CSS $|\theta
= \pi/4, \phi = 0\rangle$ (Solid line) and $|\theta = \pi/2, \phi =
0 \rangle$ (Dashed line) at different times $\Omega t = 0.75\pi/,
2.5\pi, 60.25\pi$
 (Top, middle, bottom rows respectively).  Left column: $q = 0$; right column $q \neq
0$. } \label{Pdensity}
\end{center}
\end{figure}

With a CSS input, the phase measurement gives a probability density
distribution
\begin{eqnarray}
P(\Phi)  & = &  \frac{1}{2 \pi} \left | \sum_{m_{x}, m_{z} = -j}^{j} e^{i (\Phi - 2 \Omega t ) m_{x}}  \right. \nonumber \\
& \times &  \left.  e^{i qt m_{x}^2/2} {\cal R}_{m_{z}}^{j}(\theta,
\phi)  d^{j}_{m_{z},m_{x}} (\pi/2) \right |^2 , \nonumber
\end{eqnarray}
where  $d^{j}_{m_{z},m_{x}} (\pi/2)=   _{z}\langle j,m_{z} |j,m_{x}
\rangle_{x}$ is the Wigner $d$-matrix:
\begin{eqnarray}
d^{j}_{m_{z},m_{x}} (\pi/2) & =  &  _{z}\langle j,m_{z}
| e^{-i\pi \hat{J}_{y}/2} |j,m_{x} \rangle_{z}  \nonumber \\
& = & \frac{1}{2^{m_{z}}} \left [ \frac{(j - m_{z})!(j + m_{z})!}{(j
- m_{x})!(j+ m_{x})!} \right ]^{1/2}  \nonumber \\
& \times &  P^{(m_{z}- m_{x}, m_{z} + m_{x})}_{j-m_{z}} (x = 0)
\end{eqnarray}
for $m_{z}- m_{x} > -1$ and $m_{z} + m_{x} > -1$. $P^{(\alpha,
\beta)}(x)$ denote the Jacobi polynomials. Symmetries give
$d^{j}_{m_{z},m_{x}} = (-1)^{m_{z} - m_{x}}d^{j}_{m_{x},m_{y}} =
d^{j}_{-m_{x}, -m_{z}}$. We plot the probability density in Fig.
\ref{Pdensity} at various times starting from the initial states $|
\theta = \pi/4 \rangle$ (Solid line) and $| \theta = \pi/2 \rangle$
(Dashed line). The initial Dicke state $|\theta = 0 \rangle$ is not
considered as it is orthogonal to the projective measurement on the
phase. In order to highlight the effect of nonlinearity on the
measurement of $\phi_1$, we plot in the left column the case of
$q=0$ for comparison with the corresponding results with $q \neq 0$
in the right column. We choose $q = 3/N$ which corresponds to the
Josephson regime\cite{choi2}. The presence of nonlinearity generally
degrades the performance of the interferometer as evidenced by the
increase in the width of the probability distribution. It is also
notable that the probability density becomes multiple peaked after a
long time. This may be interpreted as the generation of a
superposition state due to nonlinearity as studied by Yurke and
Stoler\cite{yurkestoler}. It is clear that to use the TBEC as an
effective interferometer based on projective measurement onto phase
states, $q$ needs to be minimzed and the time of measurement must be
kept relatively short.

\begin{figure}
\begin{center}
\centerline{\includegraphics[height=7cm]{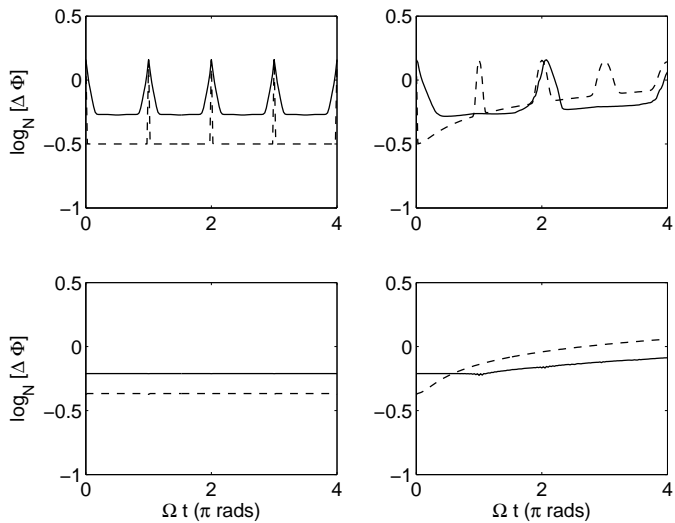}}
\caption{Time evolution of $\log_{N}[\Delta \Phi]$ for the initial
CSS $|\theta = \pi/4, \phi = 0 \rangle$ (Solid line) and $|\theta =
\pi/2, \phi = 0 \rangle$ (Dashed line).  Top row: direct calculation
from the probability density $P(\Phi)$; Bottom row: Cramers-Rao
lower bound. Left column: $q = 0$; right column $q \neq 0$.}
\label{variance}
\end{center}
\end{figure}

The  uncertainty in phase measurement can be studied using the
standard techniques of probability theory, particularly the
Cramers-Rao lower bound (CRLB). The CRLB establishes the lower bound
on the phase shift estimate where the phase uncertainty scales as
$\Delta \Phi = 1/\sqrt{F_{n}}$ where $F_{n}$ is the Fisher
information defined by
\begin{eqnarray}
F_{n} & = & \frac{1}{2 \pi} \int_{-\pi}^{\pi} \left [
\frac{d}{d\Phi} \ln P(\Phi) \right ]^{2} P(\Phi) d \Phi .
\end{eqnarray}
In Fig. \ref{variance}, we plot the quantity $\log_N [\Delta \Phi]$
where $N$ is the total number of atoms and $\Delta \Phi$ is the
uncertainty in phase. We used the standard deviation $\Delta \Phi$
calculated directly from the probability distribution $P(\Phi)$
(Fig. \ref{Pdensity}) and the CRLB, where the CRLB effectively gives
a time averaged value of the directly calculated uncertainty. It is
noted that in all these figures $\Delta \Phi \geq 1/N^{1/2}$, the
standard quantum limit. It is also noted that although the
uncertainty associated with the $| \theta = \pi/2 \rangle$ state is
lower than that of the $| \theta = \pi/4 \rangle$ state for $q = 0$
it quickly loses this advantage with $q > 0$, indicating sensitivity
to dephasing due to interatomic collisions. The $| \theta = \pi/4
\rangle$ state is therefore a more robust state for interferometry
in the presence of nonlinearity.

\begin{figure}
\begin{center}
\centerline{\includegraphics[height=7cm]{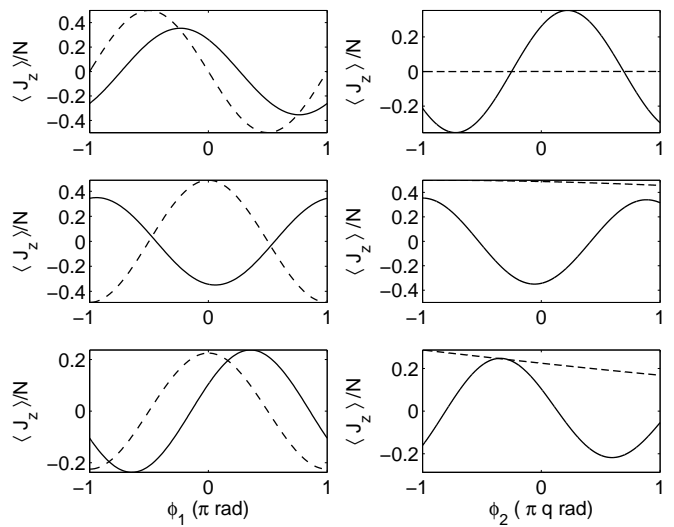}}
\caption{$\langle \hat{J}_{z} \rangle/N$ as a function of the
changes in the phase shifts $\phi_1$ (Left column) and $\phi_2$
(Right column) for the initial CSS $|\theta = 0, \phi = 0 \rangle$
(Dashed line) and $|\theta = \pi/4, \phi = 0 \rangle$ (Solid line)
at different times $\Omega t  = \pi/4, \pi, 6\pi$ (Top, middle,
bottom rows respectively).} \label{Jzplot}
\end{center}
\end{figure}

Next, instead of projective measurement onto a phase state, we
consider projective measurement of the atom number difference. The
total number of atoms measured indicates the number of ``input''
atoms while the atom number difference, $\langle
\hat{J}_{z}(\phi^{\prime}_1,\phi^{\prime}_2) \rangle \equiv \langle
\psi(0) | \hat{\cal I}^{\dagger}(\phi^{\prime}_1,\phi^{\prime}_2)
\hat{J}_{z} \hat{\cal I}(\phi^{\prime}_1,\phi^{\prime}_2)| \psi(0)
\rangle$, allows us to infer the phase shift and is equivalent to
measuring the number of atoms at each of the output port of a
typical Mach-Zehnder interferometer. An analytic expression for
$\langle \hat{J}_{z} \rangle$ is given by\cite{choi2}:
\begin{eqnarray}
\langle \hat{J}_{z} (\phi^{\prime}_1,\phi^{\prime}_2) \rangle &=&
-\sum_{m=-N/2}^{N/2-1} {\cal D}
(\theta,m)  \tan^{-1}\left(\frac{\theta -\pi /2}{2}\right ) \nonumber \\
&  \times & \cos \left[\phi^{\prime}_1 - \phi^{\prime}_2 \left( m+
\frac{1}{2}\right) \right],  \label{Jzt2}
\end{eqnarray}
where we have defined ${\cal D}(\theta,m) = C_{N/2+m+1}^{N}\left(
\frac{N}{2}+m+1 \right ) \cos ^{2N}\left( \frac{\theta -\pi /2}{2}
\right)  \tan^{N-2m}\left( \frac{\theta -\pi /2}{2}\right)$. Figure
\ref{Jzplot} shows $\langle \hat{J}_{z} \rangle/N$ as a function of
the changes in the phase shifts $\phi_1$ and $\phi_2$ for the
initial states $|\theta = 0, \phi = 0 \rangle$ and $|\theta = \pi/4
\rangle$ at different times $\Omega t  = \pi/4, \pi, 6\pi$. In
contrast to the earlier phase state projection method, the initial
CSS $|\theta = \pi/2 \rangle$ is known as a ``self-trapping'' state
in the new context of projective number measurement, and gives
trivial results. Since the interferometry is carried out at fixed
times, we see in Fig.~\ref{Jzplot} clear sinusoidal fringes, without
the ``collapses and revivals'' typical of temporal evolution. Even
when measuring $\phi_2$ we see clear fringes for a range of values
around $-3 q \ldots 3 q$ for the initial state $|\theta = \pi/4
\rangle$. This is possible because, for this choice of $\theta$, the
factor ${\cal D} (\theta = \pi/4,m)$ is narrow enough to limit the
interfering effect of summing up the cosine terms. On the other
hand, ${\cal D} (\theta = 0,m)$ is wider and the resulting
interference fringes do not allow for a sensitive detection of small
variations in $\phi_2$.

\begin{figure}
\begin{center}
\centerline{\includegraphics[height=7cm]{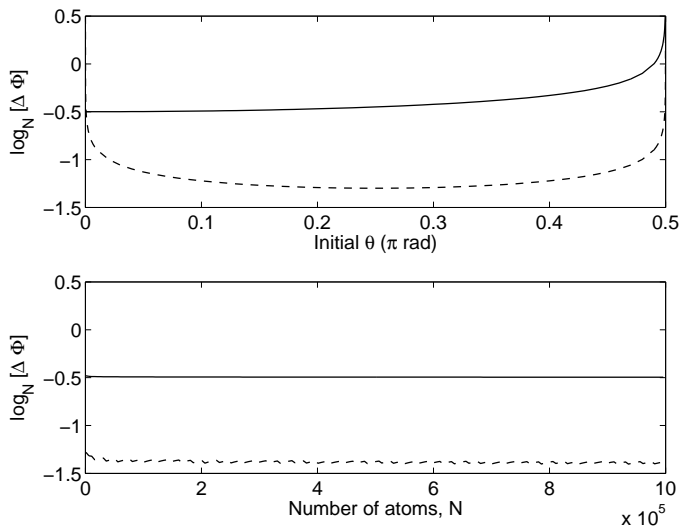}}
\caption{Top: $\log_{N}[\Delta \phi_k ]$ plotted as a function of
the angle $\theta$ of the initial CSS with $N = 1000$. Solid line:
$k = 1$, Dashed line: $k = 2$. Bottom: same quantity plotted as a
function of the number of atoms, $N$. Solid line: $k=1$ with $\theta
= 0$.  Dashed line: $k=2$ with $\theta = \pi/4$.} \label{deltaphi}
\end{center}
\end{figure}

Finally, we consider the phase resolution for this scheme which is
given by:
\begin{equation}
[\Delta \phi_k]^{2} =  \frac{[\Delta \hat{J}_{z}]^2}{|\partial
\langle \hat{J}_{z} \rangle /\partial \phi_k|^2} \;\;\;\; k = 1, 2
\label{deltaphi2}
\end{equation}
where, as found in Ref. \cite{choi2}, the variance is $[\Delta
\hat{J}_{z}]^{2} = \langle \hat{J}_{z}^{2} \rangle - \langle
\hat{J}_{z} \rangle^{2}$  with $\langle \hat{J}_{z}^{2}(t)\rangle =
\frac{1}{4} \sum_{m=-N/2}^{N/2-1} {\cal D}(\theta,m)\tan^{-2}
(\theta/2 -\pi /4) (N/2 - m )  + {\cal D}(\theta,m) ( N/2 + m + 1) +
\sum_{m=-N/2}^{N/2-2} [ \frac{1}{2} {\cal D}(\theta,m) \tan^{-2}
(\theta/2 -\pi /4)(N/2 - m - 1) ] \cos [2\phi^{\prime}_1 -
2\phi^{\prime}_2(m+1)]$. Since the denominator of Eq.
(\ref{deltaphi2}) involves a function of the form $\sin
\left[\phi^{\prime}_1 - \phi^{\prime}_2 \left( m+ \frac{1}{2}\right)
\right]$, the quantity $[\Delta \phi_k]^{2}$ is minimized for the
values of $\phi^{\prime}_1 - \phi^{\prime}_2 \left( m+ \frac{1}{2}
\right) = \pm \pi/2$. This indicates that the measurement accuracy
is dependent on the measurement values, where results such as
$\phi^{\prime}_1 = \pm \pi/2$ and $\phi^{\prime}_2 = 0$ give optimum
results. In particular, for a large number of atoms $N$ one can
approximate the coefficient ${\cal D}(\theta,m)$  by $\sqrt{N/\pi}
e^{-(2m - N \sin \theta)^2/ N }$ and replace the sums by integrals
$\int {\cal D}(\theta, x) dx = N$ and $\int N {\cal D}(\theta, x) dx
= \int  x {\cal D}(\theta, x) dx = N^{2}$. This leads to $[\Delta
\hat{J}_{z}]^{2} \sim \alpha N + \beta N^2$ and $|\partial \langle
\hat{J}_{z} \rangle /\partial \phi_1|^2 \sim \gamma N^2$ and
$|\partial \langle \hat{J}_{z} \rangle /\partial \phi_2|^2 \sim
\gamma N^4$ where $\alpha, \beta, \gamma$ are constants so that:
\begin{equation}
[\Delta \phi_k]^{2} \sim  \frac{1}{\gamma} \left [ \alpha N^{-2k +
1} + \beta  N^{-2k+2} \right ]
\end{equation}
where $k = 1, 2$. For $k = 1$ one has $\Delta \phi_1 \sim 1/N^{1/2}$
i.e. the standard quantum limit in accuracy for the measurement of
$\phi_1$. On the other hand, it is remarkable that with $k=2$ i.e.
measurement of the phase shift due to the interatomic interactions,
$\Delta \phi_2 \sim 1/N^{3/2} < 1/N$, implying that, although not
reaching the theoretical limit of $1/N^{2}$\cite{boixo}, such
measurement for this CSS input state has uncertainty below the
Heisenberg limit. We have verified this estimate numerically; the
quantity $\log_N \Delta \phi_k$ calculated as a function of the
initial angle of the CSS, $\theta$ at the optimal values of $\phi_1$
and $\phi_2$ is plotted in Fig. \ref{deltaphi}. The solid and the
dashed line represent the uncertainty in the measurement of $\phi_1$
and $\phi_2$ respectively. It is clear that the best result is
obtained for $\theta = 0$ for the measurement of $\phi_1$ and
$\theta = \pi/4$ for the measurement of $\phi_2$. In the bottom
panel, we plot the result as a function of atom numbers for these
chosen values of $\theta$. It shows that the result is independent
of number of atoms and, on average, $\Delta \phi_1 \sim 1/N^{1/2}$
and $\Delta \phi_2 \sim 1/N^{7/5}$ which is indeed very close to the
above estimate.

In summary, we have shown that a TBEC with a Josephson-like coupling
directly maps onto a nonlinear Ramsey interferometer where the phase
shifts due to linear and nonlinear variations in the Hamiltonian are
measured. The system is already experimentally available and the
state we consider is the realistic coherent spin state rather than
some exotic quantum state. It was found that projective phase
measurement reaches the standard quantum limit in accuracy while,
remarkably, projective number measurement of the phase shifts due to
interatomic interactions were found able to overcome the Heisenberg
limit, suggesting new implications for quantum metrology.

\end{document}